\documentstyle[12pt]{article}
\author 
{Wen-Xiu MA\\
 {\small Institute of Mathematics, Fudan University, 
Shanghai 200433, P. R. of China}\\
{\small Automath Institut, Universit\"at-GH Paderborn,
D-33098 Paderborn, Germany} \\{\small 
email: wenxiuma{\textrm{$@$}}uni-paderborn.de}}
\title
{Darboux Transformations for a Lax Integrable System in 
$2n$-Dimensions}
\date{\nonumber}
\begin{document}
\maketitle

\setlength{\baselineskip}{14.5pt}
\newcommand{\R}{\mbox{\rm I \hspace{-0.9em} R}}
\newcommand{\C}{\mbox{\rm l \hspace{-1.2em} C}}
\def \la {\lambda}
\def \La {\Lambda}
\def \be {\beta}
\def \al{\alpha}
\def \del{\delta}
\def \Del{\Delta}
\def \vare{\varepsilon}

\def \part {\partial}
\def \be {\begin{equation}}
\def \ee {\end{equation}}
\def \ba {\begin{array}}
\def \ea {\end{array}}
\newcommand{\eqnsection}{
   \renewcommand{\theequation}{\thesection.\arabic{equation}}
   %\makeatletter \csname $addtoreset\endcsname \makeatother
}
\eqnsection

\newtheorem{theorem}{Theorem}[section]

\begin{abstract}
A $2n$-dimensional Lax integrable system is proposed by
a set of specific spectral problems. It contains  
Takasaki equations, the self-dual Yang-Mills 
equations and its integrable hierarchy as examples.
An explicit formulation of Darboux transformations is established for 
this Lax integrable system.
The Vandermonde and 
generalized Cauchy determinant formulas lead to a description for deriving 
explicit solutions and thus
some rational  and analytic 
solutions are obtained.
\end{abstract}
%\tableofcontents
 
%\begin{center}
%Running title:

%Key words:

\section{Introduction}
\setcounter{equation}{0}
Darboux transformations provides us a purely algebraic, powerful
 method to construct
solutions for systems of nonlinear equations \cite{MatveevS}.
Its key is to expose a kind of covariant properties that the corresponding 
spectral problems 
possess. There have been many tricks to do this for getting explicit
solutions to various soliton equations including KdV equation, 
KP equation, Davey-Stewartson equation, Veselov-Novikov
equation etc.
(see for example 
\cite{Levi},\cite{AthorneN},\cite{Nimmo},\cite{LebleU},\cite{GuZ} 
and references 
therein). Darboux transformations can also be applied to generating
multi-soliton solutions to soliton equations and 
the Darboux covariance makes it   possible to  construct
a series of exactly solvable systems of supersymmetric quantum mechanics
\cite{MatveevS}.

In this letter, we would like to establish a kind of Darboux transformations
for a Lax integrable system in $2n$-dimensions which we will
introduce. The dimension reductions 
of this system contain some interesting and important equations as 
examples. Among them are Takasaki equations \cite{Takasaki},
the  self-dual Yang-Mills (SDYM)
equations, and the self-dual Yang-Mills hierarchy \cite{AblowitzCT} etc.
It is well known that the SDYM equations (even the SDYM hierarchy)
may be reduced to many integrable
 equations in $1+1$ dimensions and in $1+2$ dimensions
when certain symmetry conditions are imposed (see for instance \cite{MasonS},
\cite{AblowitzCT}). 
Therefore our Lax integrable system  also includes
a lot of integrable soliton
equations in $1+1$ dimensions and in $1+2$ dimensions.
Recently there is still an considerable interest
in the aspect of symmetry reductions of the SDYM equations
(see \cite{LegareP}, \cite{ChakravartyKN},  
a quite detailed list
of  the relevant references is included in \cite{LegareP}). 
This also increases, to a great 
extent, the validity of the Ward's conjecture \cite{Ward}:
many (and perhaps all?) integrable equations may be derived from the SDYM 
gauge field equations or its generations by reduction.

This letter is organized as follows. In Section 2, we derive a Lax 
integrable system starting from a set of specific spectral problems
and display a few concrete systems of nonlinear equations. The corresponding 
Darboux transformations are established in Section 3 and an explicit 
description of a broad class of solutions is proposed
 by means of  the resulting Darboux 
transformations. Section 4 contains some further discussions
and two remarks, along with 
a more general system whose spectral problems include negative powers of
a spectral parameter.

\section{Lax integrable system}
\setcounter{equation}{0}

Let the differential operators $L_i$, $1\le i\le n$,
 be defended by 
\be L_i=L_i(\la )=\frac {\part }{\part p_i}-a_i(\la )\frac {\part }{\part x_i}
=\frac {\part }{\part p_i}-\bigl ( \sum_{k=0}^{M}a_{ik}\la ^{k+1}\bigr)
\frac {\part }{\part x_i}, \ 1\le i\le n, \ee
where the coefficients $a_{ik}$, $0\le k\le M,\, 1\le i\le n$,
 are constants, and 
$p=(p_1,p_2,\cdots,p_n),$ $ x=(x_1,x_2,\cdots,x_n)\in \C^n$ or $\R^n$
are two vectors of 
independent variables.
We consider a set  of spectral problems
\begin{equation}
 L_i(\la )\Psi =-A_i(\la )\Psi=-\bigl (\sum_{l=0}^M A_{il}\la ^l\bigr)
\Psi,\ 1\le i\le n ,\label{laxs}
\end{equation}
where $\la $ is a spectral parameter, $\Psi$ is a $N\times N$ matrix 
of eigenfunctions, and 
$ A_{il},\ 1\le i\le n,\, 0\le l\le M$, are all $N\times N$ matrices of 
potential functions depending on $p,x$.
This spectral problem contains Takasaki case \cite{Takasaki}: 
$M=0,\ 
a_{i}(\la )=\la ,\ 1\le i\le n;$ 
 Gu and Zhou case \cite{GuZ}: $a_i(\la )=0,\
1\le i\le n,\ A_i(\la )=\la J_i+P_i, \ 1\le i\le n-1$, where $J_i,\,P_i$ are
the matrices given in \cite{GuZ};
and Gu case \cite{Gu}: $a_{i}(\la )=\la 
,\ 1\le i\le n-1,\ a_{n}(\la )=0,\ A_{i}(\la )=A_{i0},\ 
1\le i\le n-1$. We assume that $n,N\ge 2$ in order to obtain nonlinear 
integrable systems (we shall explain this later). 
Noting that $L_i,\ 1\le i\le n,$ are all 
linear operator, we can calculate that 
\begin{eqnarray}&& \quad
 L_j(\la )L_i(\la )\Psi=L_j(\la )(-\sum_{l=0}^M\la ^lA_{il}\Psi) \nonumber\\&&
=-\sum_{l=0}^M\la ^lA_{il}L_j(\la ) \Psi-\sum_{l=0}^M\la ^l(L_j(\la )
A_{il})\Psi
\nonumber\\&&
=\sum_{l=0}^M\la ^lA_{il}\sum_{k=0}^M\la ^kA_{jk}\Psi-
\sum_{l=0}^M\la ^l(\frac {\part A_{il}}{\part p_j}-a_i(\la )\frac 
{\part A_{il}}
{\part x_j})\Psi\nonumber\\&&
=\sum_{m=0}^{2M}\la ^m\Bigr(\sum_{ \ba {c} \vspace{-7mm}\\
  {\scriptstyle   k+l=m}
\vspace{-2mm}\\
{\scriptstyle  0\le k,l\le M}\ea }
A_{ik}A_{jl}\Bigl)\Psi -\sum_{m=0}^M\la ^m \frac {\part A_{im}}{\part p_j}\Psi 
+\sum_{m=0}^{2M}\la ^{m+1}
\Bigr(\sum_{ \ba {c}\vspace{-7mm}\\{\scriptstyle  k+l=m}\vspace{-2mm}\\ 
{\scriptstyle  0\le k,l\le M}
\ea }
a_{jk}\frac {\part A_{il}}{\part x_j}\Bigl)\Psi.\nonumber \end{eqnarray}
Therefore we see that the compatibility conditions \be 
L_j(\la )L_i(\la )\Psi=L_i(\la )L_j(\la )\Psi,
\ 1\le i,j\le n,\label{ccondtions}\ee
are expressed as 
\begin{eqnarray} &&\sum_{m=0}^{2M}\la ^m\sum_{ 
 \ba {c}\vspace{-7mm}\\  {\scriptstyle   k+l=m}
\vspace{-2mm}\\
{\scriptstyle  0\le k,l\le M}\ea }
A_{ik}A_{jl} -\sum_{m=0}^M\la ^m \frac {\part A_{im}}{\part p_j}
+\sum_{m=0}^{2M}\la ^{m+1}
\sum_{\vspace{-1mm} \ba {c}\vspace{-7mm}\\ {\scriptstyle  k+l=m}\vspace{-2mm}\\ 
{\scriptstyle  0\le k,l\le M}
\ea }
a_{jk}\frac {\part A_{il}}{\part x_j}\nonumber \\ &=&
\sum_{m=0}^{2M}\la ^m\sum_{  \ba {c}\vspace{-7mm}\\  {\scriptstyle   k+l=m}
\vspace{-2mm}\\
{\scriptstyle  0\le k,l\le M}\ea }
A_{jk}A_{il} -\sum_{m=0}^M\la ^m \frac {\part A_{jm}}{\part p_i}
+\sum_{m=0}^{2M}\la ^{m+1}
\sum_{ \ba {c}\vspace{-7mm}\\ {\scriptstyle  k+l=m}\vspace{-2mm}\\ 
{\scriptstyle  0\le k,l\le M}
\ea }
a_{ik}\frac {\part A_{jl}}{\part x_i}.\quad
  \label{polyoflambda}\end{eqnarray}
Equating coefficients of the terms $\la ^0$;
$\la ^m, \ 1\le m\le M$; $\la ^m,\ M+1\le m\le 2M$ and  $\la ^{2M+1}$ in 
the above equation 
 lead  to the following Lax integrable system in
$2n$-dimensions:
\begin{eqnarray}&& [A_{i0},A_{j0}]-\frac {\part A_{i0}}{\part p_j}+
\frac {\part A_{j0}}{\part p_i}=0,\ 1\le i,j\le n,\label{laxis1}\\&
&\sum_{\ba {c}\vspace{-7mm}\\  {\scriptstyle k+l=m}\vspace{-2mm}\\ 
{\scriptstyle 0\le k,l\le M}\ea}[A_{ik},A_{jl}]-
\frac {\part A_{im}}{\part p_j}+
\frac {\part A_{jm}}{\part p_i}+
\sum_{\ba {c}\vspace{-7mm}\\  {\scriptstyle k+l=m-1}\vspace{-2mm}\\ 
{\scriptstyle 0\le k,l\le M}\ea}\Bigr(a_{jk}\frac {\part A_{i
l}}{\part x_j}-a_{ik}
\frac {\part A_{jl}}{\part x_i} \Bigl)=0, \quad\nonumber\\&&
\qquad\qquad\qquad\qquad\qquad\qquad\qquad\qquad\quad    
1\le m\le M,\ 1\le i,j\le n,\label{laxis2}\\
&&\sum_{\ba {c}\vspace{-7mm}\\  {\scriptstyle k+l=m}\vspace{-2mm}\\ 
{\scriptstyle 0\le k,l\le M}\ea}[A_{ik},A_{jl}]\
+\sum_{\ba {c}\vspace{-7mm}\\  {\scriptstyle k+l=m-1}\vspace{-2mm}\\ 
{\scriptstyle 0\le k,l\le M}\ea}\Bigr(a_{jk}\frac {\part A_{i
l}}{\part x_j}-a_{ik}
\frac {\part A_{jl}}{\part x_i} \Bigl)=0, \nonumber\\&&
\qquad\qquad\qquad\qquad\qquad\qquad\qquad \qquad\quad 
M+1\le m\le 2M,\ 1\le i,j\le n,\label{laxis3}\\&&
  a_{jM}\frac {\part A_{i
M}}{\part x_j}-a_{iM}
\frac {\part A_{jM}}{\part x_i} =0,\ 1\le i,j\le n,\label{laxis4}
\end{eqnarray}
where $[\cdot, \cdot ]$ denotes Lie bracket of matrix Lie algebras.
Evidently for the case of $N=1$, 
the commutators of the matrices $A_{ik},\,A_{jl}$
are all equal to zero and thus the above system is simplified into a linear 
system, which isn't what we need. Besides, if $n=1$, the above system 
holds automatically and thus doesn't need any consideration.
The Lax integrable system consisting of (\ref{laxis1}), (\ref{laxis2}),
(\ref{laxis3}) and (\ref{laxis4})
 includes some interesting and important systems of 
equations as examples. For instance,
we have a few systems of equations as follows.

\noindent 
{\it Example 1}: $M=0,\ a_{i}(\la )=1,\ 1\le i\le n$, 
corresponds to Takasaki case \cite{Takasaki}
which gives rise to the equations
\begin{eqnarray}&&
[A_{i0},A_{j0}]-\frac {\part A_{i0}}{\part p_j}+
\frac {\part A_{j0}}{\part p_i}=0,\ 1\le i,j\le n,\label{takasakis1}\\&&
\frac {\part A_{i0}}{\part x_j}-\frac {\part A_{j0}}{\part x_i}=0,\ 
1\le i,j\le n.\label{takasakis2}
\end{eqnarray}
This system may be changed into
\be \frac {\part }{\part x_i}\Bigr(\frac {\part J}{\part p_j}J^{-1}\Bigl)
- \frac {\part }{\part x_j}\Bigr(\frac {\part J}{\part p_i}J^{-1}\Bigl)=0
,\ 1\le i,j\le n\ee
or 
\be \frac {\part }{\part x_i}\Bigr(J^{-1}\frac {\part J}{\part p_j}\Bigl)
- \frac {\part }{\part x_j}\Bigr(J^{-1}\frac {\part J}{\part p_i}\Bigl)=0
,\ 1\le i,j\le n\label{gyang}\ee
after making a transformation 
\[ A_{i0}=-\frac {\part J}{\part p_i}J^{-1}\ \ \textrm{or}\ \ A_{i0}=J^{-1}
\frac {\part J}{\part p_i},\ 1\le i\le n,\]
respectively.
The equations (\ref{gyang}) with $n=2$ is  
a kind of version of the anti-self-dual 
Yang-Mills equations due to Pohlmeyer
\cite{Pohlmeyer}. Its inverse scattering has been analyzed by Beals and 
Coifman \cite{BealsC} and it
may be rewritten in the original Yang's equation
``in $R$-gauge'' \cite{Yang} upon choosing
\[J=\frac1 u \left[\ba {cc} 1 & w\\ v& u^2+vw\ea \right].\]

\noindent 
{\it Example 2}: Let $M=1,\ n=2,\ a_1(\la )=-\la ,\
 a_2(\la )=\la $. We obtain the 
equations
\begin{eqnarray}&&[A_{10},A_{20}]-\frac {\part A_{10}}{\part p_2}+
\frac {\part A_{20}}{\part p_1}=0,\label{sdym1}
\\&& 
[A_{10},A_{21}]+[A_{11},A_{20}]
-\frac {\part A_{11}}{\part p_2}+
\frac {\part A_{21}}{\part p_1}+\frac {\part A_{10}}{\part x_2}+
\frac {\part A_{20}}{\part x_1}=0,\label{sdym2}
\\&& 
[A_{11},A_{21}]+\frac {\part A_{11}}{\part 
x_2}+
\frac {\part A_{21}}{\part x_1}=0,\label{sdym3}
\end{eqnarray}
which yields the SDYM equations discussed in Ref. \cite{AblowitzC},
upon making $A_{21}\to -A_{21}$.
 Here $A_{10},A_{11},A_{20},A_{21}$ are all the  Yang-Mills potentials.
The corresponding inverse scattering 
problem was introduced by Belavin and Zakharov
 \cite{BelavinZ} many years ago. Recently a scheme of symmetry reduction
of the Lax pairs for the SDYM equations with respect 
to an arbitrary subgroup of their conformal group has been described by 
Legar\`e and Popov \cite{LegareP}
and accordingly
the compatibility conditions of the reduced Lax pairs lead
 to the SDYM equations reduced
under the same symmetry group.

\noindent 
{\it Example 3}: The case of  $M=1, a_{i}(\la )=\la ,\ 1\le i\le n$,
yields  the equations
\begin{eqnarray}&&[A_{i0},A_{j0}]-\frac {\part A_{i0}}{\part p_j}+
\frac {\part A_{j0}}{\part p_i}=0,\ 1\le i,j\le n,
\\&& 
[A_{i0},A_{j1}]+[A_{i1},A_{j0}]-\frac {\part A_{i1}}{\part p_j}+
\frac {\part A_{j1}}{\part p_i}
+
\frac {\part A_{i0}}{\part x_j}-
\frac {\part A_{j0}}{\part x_i}=0,
\ 1\le i,j\le n,\qquad
\\&& 
[A_{i1},A_{j1}]+\frac {\part A_{i1}}{\part 
x_j}-
\frac {\part A_{j1}}{\part x_i}=0,\ 1\le i,j\le n.
\end{eqnarray}
In comparison
 with the SDYM equations (\ref{sdym1}),(\ref{sdym2}),(\ref{sdym3}),
these equations may be referred to  as the generalized SDYM equations.

\noindent 
{\it Example 4}:
 Let $M=1, a_{i}(\la )=\la ^2,\ 1\le i\le n$. We obtain the equations
 \begin{eqnarray}&&[A_{i0},A_{j0}]-\frac {\part A_{i0}}{\part p_j}+
\frac {\part A_{j0}}{\part p_i}=0,\ 1\le i,j\le n,
\\&& 
[A_{i0},A_{j1}]+[A_{i1},A_{j0}]
-\frac {\part A_{i1}}{\part p_j}+
\frac {\part A_{j1}}{\part p_i}=0,
\ 1\le i,j\le n,
\\&& 
[A_{i1},A_{j1}]+\frac {\part A_{i0}}{\part 
x_j}-
\frac {\part A_{j0}}{\part x_i}=0,\ 1\le i,j\le n,
\\&& 
\frac {\part A_{i1}}{\part x_j}-\frac {\part A_{j1}}{\part x_i}=0,\ 
1\le i,j\le n.
\end{eqnarray}

\noindent 
{\it Example 5}:
 The case of  $M=1, a_{i}(\la )=\la +\la ^2
,\ 1\le i\le n$, leads to the equations
\begin{eqnarray}&&[A_{i0},A_{j0}]-\frac {\part A_{i0}}{\part p_j}+
\frac {\part A_{j0}}{\part p_i}=0,\ 1\le i,j\le n,
\\&& 
[A_{i0},A_{j1}]+[A_{i1},A_{j0}]-[A_{i1},A_{j1}]
-\frac {\part A_{i1}}{\part p_j}+
\frac {\part A_{j1}}{\part p_i}=0,
\ 1\le i,j\le n,
\\&& 
[A_{i1},A_{j1}]+\frac {\part A_{i0}}{\part 
x_j}-
\frac {\part A_{j0}}{\part x_i}=0,\ 1\le i,j\le n,
\\&& 
\frac {\part A_{i1}}{\part x_j}-\frac {\part A_{j1}}{\part x_i}=0,\ 
1\le i,j\le n.
\end{eqnarray}
The systems in Examples 4,5 are new to our knowledge.
Obviously each system of
Examples 3,4,5 is a generalization 
of  Takasaki system of (\ref{takasakis1}),
(\ref{takasakis2}), since all corresponding  reductions of potentials $A_{i1}=0
,\ 1\le i\le n$,
lead to Takasaki system. 

\noindent 
{\it Example 6}:
 Let $M\ge 2,\ n=2,\ a_1(\la )=-\la ,\ a_2(\la )=\la ^{M},\ 
A_1(\la )=A_{10}+A_{11}\la  $.
 We obtain all equations but the SDYM equations (Example 2)
in the SDYM integrable hierarchy \cite{AblowitzC} 
\begin{eqnarray}&&[A_{10},A_{20}]-\frac {\part A_{10}}{\part p_2}+
\frac {\part A_{20}}{\part p_1}=0,
\\&& 
[A_{10},A_{21}]+[A_{11},A_{20}]
-\frac {\part A_{11}}{\part p_2}+
\frac {\part A_{21}}{\part p_1}+
\frac {\part A_{20}}{\part x_1}=0,
\\&& 
[A_{10},A_{2m}]+[A_{11},A_{2,m-1}]
+
\frac {\part A_{2m}}{\part p_1}+
\frac {\part A_{2,m-1}}{\part x_1}=0,\ 2\le m\le M-1,\label{yangh3}\\&& 
[A_{10},A_{2M}]+[A_{11},A_{2,M-1}]
+\frac {\part A_{2M}}{\part p_1}+
\frac {\part A_{2,M-1}}{\part x_1}+\frac {\part A_{10}}{\part x_2}=0,
\\&& 
[A_{11},A_{2M}]+\frac {\part A_{11}}{\part 
x_2}+
\frac {\part A_{2M}}{\part x_1}=0.
\end{eqnarray}
Note that when $M=2$, the equation (\ref{yangh3}) doesn't appear.
Recursion operators, symmetries and conservation laws associated with
this hierarchy have been considered by Ablowitz et al \cite{AblowitzCT}.
In fact, it follows from (\ref{yangh3}) that we have a recursion relation
\[A_{2m}=-\Bigl(\textrm{ad}_{A_{10}}+\frac {\part }{\part p_1}\Bigr)^{-1}
\Bigl(\textrm{ad}_{A_{11}}+\frac {\part }{\part x_1}\Bigr)A_{2,m-1}=R 
A_{2,m-1},\]
where $R$ is exactly a recursion operator of the SDYM equations.
The above 4-dimensional SDYM hierarchy can also be considered 
as a ``universal''
integrable hierarchy. Upon appropriate reduction and suitable choice of gauge 
group, it can produce virtually all well known hierarchy of soliton equations 
in $1+1$ or $1+2$ dimensions \cite{AblowitzCT}. 

\section{Darboux transformations}
\setcounter{equation}{0}

We recall the spectral problems (\ref{laxs}):
\[ L_i(\la )\Psi =\Bigl(\frac {\part }{\part p_i}-a_i(\la )
\frac {\part }{\part x_i}\Bigr)\Psi =-A_i(\la )\Psi ,\ 1\le i\le n.\]
Darboux transformation problem means how  we can obtain a new set of 
\be \tilde {\Psi },\ \tilde{A}_{i}(\la )=\sum_{l=0}^M\tilde A _{il}\la ^l,\ 
1\le i\le n,\ee
from a old set of 
\[\Psi , A_{i}(\la )=\sum_{l=0}^M A _{il}\la ^l,\ 1\le i\le n,\]
so that the spectral problems (\ref{laxs}) hold covariantly.
The purpose of this section is just
to give an answer to this problem.
We shall prove that for certain $\al $ to be determined,
\be \tilde \Psi =(\la I+\al )\Psi,\ I=\textrm{diag}(\,\underbrace{1,1,\cdots,1}
_{N}\,)\label{newalpha}\ee
will give a practicable new matrix of eigenfunctions. Of course
this matrix $ \al  $  has to satisfy some conditions. 

For the time being, let us observe 
what kind of conditions they should be.
Because we have 
\begin{eqnarray}&& L_i(\al )\tilde \Psi=L_i(\al ) [(\la I+\al )\Psi]
=(L_i(\al )\al )\Psi +(\la I+\al )L_i(\al )\Psi \nonumber 
\\&&
 =(L_i(\al )\al )\Psi -(\la I+\al )A_i(\la )\Psi\nonumber \\ && 
 =\Bigr(\frac{\part \al }{\part p_i}-\sum_{k=0}^{M}a_{ik}\la ^{k +1}
\frac{\part \al }{\part x_i}\Bigl )\Psi -(\la I+\al )\Bigr(\sum_{l=0}^M
A_{il}\la ^l
\Bigl )\Psi,\ 1\le i\le n,\nonumber\end{eqnarray}
and \[
-\tilde A_{i}(\la )\tilde \Psi =-\Bigr(\sum_{l=0}^M
\tilde A_{il}\la ^l \Bigl )(\la I+\al )\Psi, \ 1\le i\le n,
\]
a balance of the coefficients of the powers
 $\la ^0$; $\la ^m,\ 1\le m\le M$ and
$\la ^{M+1}$ of $L_i(\la )\tilde \Psi =-\tilde A_{i}(\la )\tilde \Psi,\ 
1\le i\le n$, yields that for $1\le i\le n$,
\begin{eqnarray}&& \frac {\part \al }{\part p_i}-\al A_{i0}=-\tilde A_{i0}\al ,
\nonumber \\&&
-a_{i,m-1}\frac {\part \al }{\part x_i}-A_{i,m-1}-\al A_{im}=-\tilde A_{i,m-1}-
\tilde A_{im}\al ,\ 1\le m\le M,\nonumber\\ &&
-a_{iM}\frac {\part \al }{\part x_i}-A_{iM}=-\tilde A_{iM}.
\nonumber \end{eqnarray}
From these equalities we get the condition for $\al $
\be \frac {\part \al }{\part p_i}=\al A_{i0}-\tilde A_{i0}\al ,\ 1\le i\le n ,
\label{oalcond} \ee
and at the same time
 a recursion formula to determine
$\tilde A_{im}$: 
\be \left \{\ba {l}\tilde A_{iM}=A_{iM}+a_{iM}\frac {\part \al }{\part x_i},
\ 1\le i\le n,\vspace{2mm}\\
\tilde A_{i,m-1}=A_{i,m-1}+a_{i,m-1}
\frac {\part \al }{\part x_i}+\al A_{im}-\tilde
A_{im}\al ,\ 1\le m\le M,\ 
\ 1\le i\le n.
\ea\right.\label{reofal}\ee
We observe (\ref{reofal}) and (\ref{oalcond}) a little more.
Notice that 
for $1\le r\le M,\ 1\le i\le n$, we may make a further calculation 
\begin{eqnarray} &&
\tilde A_{i,r-1}=A_{i,r-1}+a_{i,r-1}\frac {\part \al }{\part x_i}+\al A_{ir}
+\tilde A_{ir}(-\al )\nonumber\\&&
=A_{i,r-1}+a_{i,r-1}\frac {\part \al }{\part x_i}+\al A_{ir}
+ \Bigr(A_{ir}+a_{ir}\frac {\part \al }{\part x_i}+\al A_{i,r+1} \Bigl)(-\al )
+\tilde A_{i,r+1}(-\al )^2\qquad\quad
 \nonumber\\ &&
\cdots \cdots\nonumber\\&&
=\sum_{k=0}^{M-r}\Bigr(A_{i,r+k-1}+a_{i,r+k-1}\frac {\part \al }{\part x_i}+
\al A_{i,r+k}
\Bigl)(-\al )^{k}+\tilde A_{iM}(-\al )^{M-r+1}.
\label{dtpold} \end{eqnarray}
Therefore (\ref{reofal}) becomes 
\be  
\left\{ \ba{l}\tilde {A}_{i,r-1}=\sum_{k=0}^{M-r+1}\Bigl(A_{i,r+k-1}
+a_{i,r+k-1}
\frac {\part \al }{\part x_i} \Bigr)(-\al ) ^k \vspace{1.8mm}\\
\qquad \qquad +
\sum_{k=0}^{M-r}\al A_{i,r+k}(-\al )^k, \ 1\le r\le M,\ 1\le i\le n,
\vspace{2mm}\\
\tilde {A}_{iM}=A_{iM}+a_{iM}\frac {\part \al }{\part x_i},\ 1\le i\le n,
\ea\right.
\label{dtpo}\ee
and further the condition (\ref{oalcond}) reads as 
\begin{eqnarray}&& \frac {\part \al }{\part p_i}=
\al A_{i0}-\tilde A_{i0}\al \nonumber\\
&&=\al A_{i0}+\sum_{k=0}^MA_{ik}(-\al )^{k+1}+\sum_{k=0}^{M}a_{ik}
\frac {\part \al }{\part x_i}(-\al )^{k+1}+\sum_{k=1}^M\al A_{ik}(-\al )^k\nonumber
\\&&
=\sum_{k=0}^MA_{ik}(-\al )^{k+1}+\sum_{k=0}^{M}a_{ik}
\frac {\part \al }{\part x_i}(-\al )^{k+1}+\sum_{k=0}^M\al A_{ik}(-\al )^k
,\ 1\le i\le n.\quad\label{alcond}
\end{eqnarray}
This is a final condition to restrict the matrix $\al $ in the construction 
of new matrices of eigenfunctions according to  (\ref{newalpha}).
The following result can provide us such a kind of useful matrices.
\begin{theorem} Let $h_s,\ 1\le s\le N,$ be $N$-dimensional 
 column eigenvectors
corresponding to the spectral parameters $\la _1,\la _2,\cdots ,\la _N$,
 respectively, that is to say that the $N$-dimensional 
 column eigenvectors $h_s,\ 1\le s\le N,$ satisfy 
\be  \frac {\part h_s}{\part p_i}=a_i(\la _s) \frac {\part h_s}{\part x_i}
-A_i(\la _s)h_s, \ 1\le s\le N,\ 1\le i\le n.\label{h_scondtion}\ee
Assume that the determinant of the matrix 
$ H=[h_1,h_2,\cdots,h_N]$ is non-zero. 
Then the matrix defined by 
\be
\al =-H\Lambda H^{-1}, \ \Lambda= {diag}(\la _1,\la _2,\cdots,\la _N),
\label{alpha}\ee
satisfies the condition (\ref{alcond}). Therefore 
we have a Darboux transformation 
$ \tilde \Psi =(\la I+\al )\Psi$ and a new solution 
$ \tilde A_{il}, \ 1\le i\le n,\ 0\le l\le M
,$ defined by 
(\ref{dtpo}).\label{dttheorem}
\end{theorem}
{\bf Proof:} Noting that
\[ \frac {\part H^{-1}}{\part y_i}=-H^{-1}\frac {\part H}{\part y_i}H^{-1},\ 
y_i=p_i \ \textrm{or}\ x_i,\ 1\le i\le n, \]
we have for any $1\le i\le n$
\begin{eqnarray}&&\frac {\part \al }{\part p_i}=-\frac {\part H}{\part p_i}
\Lambda H^{-1}+H\Lambda H^{-1}\frac {\part H}{\part p_i}H^{-1},\nonumber\\&&
\frac {\part \al }{\part x_i}=-\frac {\part H}{\part x_i}
\Lambda H^{-1}+H\Lambda H^{-1}\frac {\part H}{\part x_i}H^{-1}.\nonumber
\end{eqnarray}
On the other hand, from (\ref{h_scondtion}) we obtain
\begin{eqnarray}&&
\frac {\part H}{\part p_i}=\sum_{k=0}^{M}a_{ik}
\frac {\part H}{\part x_i}
\Lambda ^{k+1}-[A_i(\la _1)h_1,\cdots, A_i(\la _N)h_N]
\nonumber\\&&
=\sum_{k=0}^{M}a_{ik}\frac {\part H}{\part x_i}
\Lambda ^{k+1}-\sum_{l=0}^MA_{il}H\Lambda ^l,\ 1\le i\le n. 
\nonumber\end{eqnarray}
In this way, we can compute that
\begin{eqnarray}&&
\frac {\part \al }{\part p_i}=-\frac {\part H}{\part p_i}
\Lambda H^{-1}+H\Lambda H^{-1}\frac {\part H}{\part p_i}H^{-1}\nonumber\\
&& =-\sum_{k=0}^{M}a_{ik}\frac {\part H}{\part x_i}
\Lambda ^{k+2}H^{-1}+\sum_{l=0}^MA_{il}H\Lambda ^{l+1}H^{-1}\nonumber\\&&
\quad +\sum_{k=0}^{M}a_{ik}H\Lambda H^{-1}\frac {\part H}{\part x_i}
\Lambda ^{k+1}H^{-1}-H\Lambda H^{-1}\sum_{l=0}^MA_{il}H\Lambda ^lH^{-1}
\nonumber\\&&
=-\sum_{k=0}^{M}a_{ik}\frac {\part H}{\part x_i}H^{-1}(-\al )^{k+2}
+\sum_{l=0}^MA_{il}(-\al )^{l+1}\nonumber\\&& \qquad
+\sum_{k=0}^{M}a_{ik}H\Lambda H^{-1}\frac {\part H}{\part x_i}
H^{-1}\al ^{k+1}+\al \sum_{l=0}^MA_{il}(-\al )^l\nonumber\\&&
=\sum_{k=0}^{M}a_{ik}\Bigl(-\frac {\part H}{\part x_i}\Lambda H^{-1}+
H\Lambda H^{-1}\frac {\part H}{\part x_i}H^{-1}\Bigr)
(-\al )^{k+1}\nonumber\\&&\qquad +\sum_{l=0}^MA_{il}(-\al )^{l+1}+\al 
\sum_{l=0}^MA_{il}(-\al )^l\nonumber\\&&
=\sum_{k=0}^{M}a_k\frac {\part \al }{\part x_i}(-\al )^{k+1}
+\sum_{l=0}^MA_{il}(-\al )^{l+1}+\al 
\sum_{l=0}^MA_{il}(-\al )^l.\nonumber
\end{eqnarray}
This shows that the matrix $\al $ defined by (\ref{alpha})
satisfies the condition (\ref{alcond}), indeed.
The rest of the proof is evident.  The proof is completed.
$\vrule width 1mm height 3mm depth 0mm$

We mention that we need a requirement that at least 
there are two different parameters 
in the set $\{\la_1,\la _2,\cdots, \la _N\}$ while making Darboux 
transformation determined by Theorem \ref{dttheorem}. Otherwise we only get 
an original solution not a new solution since $\al $ becomes a unit matrix
up to a constant factor. 
The equation (\ref{h_scondtion}) for $h_s$ is linear and hence
there is no problem to solve it. 
So Darboux transformations (\ref{newalpha}) can always engender new explicit 
solutions once one solution is found.  
Furthermore from   the  solution induced by
 Darboux transformation, we may make Darboux 
transformation once more and obtain
another new solution. This process can be done continually and usually it
may yield a series of multi-soliton solutions \cite{GuZ}, \cite{Gu}.

\section{Rational  and analytic solutions}
\setcounter{equation}{0}

We make Darboux transformation 
starting from a special initial solution: zero solution $A_{il}=0, \ 1\le i\le n,\ 0\le j\le M.$ Let $\la _1,\la _2,\cdots ,\la _N$ be arbitrary parameter
so that at least two of them are different. In this way, 
we obtain a new solution
\be  
\left\{ \ba {l}\tilde {A}_{i,r-1}=\sum_{k=0}^{M-r+1}a_{i,r+k-1}
\frac {\part \al }{\part x_i} (-\al ) ^k 
, \ 1\le r\le M,\ 1\le i\le n,
\vspace{2mm}\\
\tilde {A}_{iM}=a_{iM}\frac {\part \al }{\part x_i},\ 1\le i\le n,
\ea \right.
\label{dtp}\ee
where the matrix $\al $ is defined (\ref{alpha}). 

Evidently it is a crucial
point to construct an invertible matrix $H=[h_1,h_2,\cdots ,h_N]$.
We shall utilize two special determinants to generate this kind of matrices
that we need. 
The one is the Vandermonde determinant formula
\be \left | \ba {cccc}1&1&\cdots &1\\
a_1&a_2&\cdots &a_N\\
\vdots &\vdots &\ddots &\vdots \\
a_1^{N-1}&a_2^{N-1}&\cdots &a_N^{N-1}\ea \right|=\prod _{i>j}(a_i-a_j), \ee
and the other 
one is a generalized Cauchy determinant formula developed 
by Constantinescu \cite{FC}
\be (-1)^{N(N-1)/2}\left | \ba {ccc}\Delta _{11}&\cdots &\Delta _{1n_2}\\
\vdots & \ddots &\vdots \\
\Delta _{n_11}&\cdots &\Delta _{n_1n_2}
\ea \right|=\frac {\prod _{1\le i<j\le N_1}(a_i-a_j)^{n_1^2}
\prod _{1\le i<j\le N_2}(b_i-b_j)^{n_2^2}}{\prod _{1\le i\le N_1,\,
1\le j\le N_2}(a_i-b_j)^{n_1n_2}} , \ee
where $n_1N_1=n_2N_2=N$ and 
the matrices $\Delta _{kl}, \ 1\le k\le n_1,\ 1\le l\le n_2,$ are  defined by
\be \Delta _{kl}={{k+l-2}\choose {k-1}}\left[
\ba {ccc}\frac 1{(a_1-b_1)^{k+l-1}}&\cdots &\frac 1{(a_1-b_{N_2})^{k+l-1}}\\
\vdots & \ddots &\vdots \\
\frac 1{(a_{N_1}-b_1)^{k+l-1}}&\cdots &\frac 1{(a_{N_1}-b_{N_2})^{k+l-1}}
\ea\right].\ee
When $N_1=N_2=N$ (this moment $n_1=n_2=1$), 
the above generalized Cauchy determinant formula is reduced to
the Cauchy determinant formula.

Let $i_1,i_2,\cdots, i_N$ be $N$ integers and $n_1N_1=n_2N_2=N$.
 For simplicity, we accept 
\be a(\la _i)\cdot p +x=(a_1(\la _i)p_1+x_1,\cdots,a_n(\la _i)p_n+x_n),\ 
1\le i\le N.\ee 
We choose the first class of the matrices $H$ as follows
\be H_1=\left[\ba {cccc} f_1^{i_1}&f_2^{i_2}&\cdots &f_N^{i_N}\\
f_1^{i_1+1}&f_2^{i_2+1}&\cdots &f_N^{i_N+1}\\
\vdots &\vdots &\ddots &\vdots\\
f_1^{i_1+N-1}&f_2^{i_2+N-1}&\cdots &f_N^{i_N+N-1}
\ea \right], \label{h1} \ee 
with the functions $ f_i=f_i(a(\la _i)\cdot p +x),\ 1\le i\le N$.
 This is a little more general
Vandermonde matrix and hence we have
\be  \textrm{det}H_1=f_1^{i_1}f_2^{i_2}\cdots f_N^{i_N}\prod _{i>j}(f_i-f_j).
\label{deth1}\ee 
We obtain 
$\al _1=-H_1\Lambda _1H_1^{-1}$, where $\Lambda _1=\textrm{diag}(\la _1,\la _2,
\cdots, \la _N)$, provided that det$H_1\ne 0$.
 We choose the second class of the matrices $H$ as follows
\be H_2=\left[\ba {ccc} \Delta _{11}(f,g,\mu )&\cdots &\Delta _{1n_2}
(f,g,\mu )\\
\vdots  &\ddots &\vdots\\ \Delta _{1n_1}(f,g,\mu )&\cdots &\Delta _{n_1n_2}
(f,g,\mu )
\ea \right], \label{h2} \ee 
where $\Delta _{kl}(f,g,\mu ),\ 1\le k\le n_1,\ 1\le l\le n_2,$ 
are given by
\be \Delta _{kl}(f,g,\mu )={{k+l-2}\choose {k-1}}\left[\ba {ccc}
\frac{f_1^{i_{(l-1)N_2+1}}}{(\mu _{1}-g_1)^{k+l-1}}&\cdots& 
\frac{f_{N_2}^{i_{lN_2}}}{(\mu _1-g_{N_2})^{k+l-1}}\\
\vdots  &\ddots &\vdots\\
\frac{f_1^{i_{(l-1)N_2+1}}}{(\mu _{N_1}-g_1)^{k+l-1}}&\cdots& 
\frac{f_{N_2}^{i_{lN_2}}}{(\mu _{N_1}-g_{N_2})^{k+l-1}}
 \ea \right]\ee
with the constants $\mu _i,\ 
1\le i\le N_1,$ and the functions 
$ f_i=f_i(a(\la _i)\cdot p +x)$,
 $ g_i=g_i(a(\la _i)\cdot p +x),\ 1\le i\le N_2.$ 
The determinant of this matrix $H_2$ can be computed by
the generalized Cauchy determinant formula and eventually we obtain
\be \textrm{det}H_2=(-1)^{N(N-1)/2}f_1^{\sum_{l=1}^{n_2}i_{(l-1)N_2+1}}\cdots
 f_{N_2}^{\sum_{l=1}^{n_2}i_{lN_2}}\frac {\prod_{1\le i<j\le N_1}
(\mu _i-\mu _j)^{n_1^2}
\prod_{1\le i<j\le N_2}(g _i-g _j)^{n_2^2}}{\prod_{1\le i\le N_1,1\le j\le N_2}
(\mu _i-g_j)^{n_1n_2}}.\label{deth2}\ee
In this way, we have 
$\al _2=-H_2\Lambda _2H_2^{-1}$, where $\Lambda _2=\textrm{diag}(
 \la _1,
\underbrace{\cdots, \la _{N_2};\cdots;\la _1,
\cdots}_{n_2}, \la _{N_2})$ provided that det$H_2\ne 0$.

In what follows, we restrict our consideration within the real field but the 
case of the complex field is completely similar.

{\bf (1) Rational  function solutions:}

\noindent (1.1) Let $f_1,\cdots ,f_N: \R ^n\to \R$ be non-zero distinct 
rational  functions.
At this moment, we have $f_i(a(\la  
_i)\cdot p +x)\mbox{ $ / \hspace{-0.9em}\equiv $ }0,\ 1\le i\le N,$ and $f_i(a(\la  
_i)\cdot p +x)\mbox{ $ / \hspace{-0.9em}\equiv $ }  f_j(a(\la  _j)\cdot p +x),\ i\ne j$. 
Otherwise we have $f_i(x)\equiv 0, \ 1\le i\le N,$ and 
 $f_i(x)\equiv f_j(x),\ i\ne j$, by setting  $p=0$,
which contradict to the original hypothesis.
It follows from (\ref{deth1}) that in this case, 
 $\textrm{det}H_1$ 
is a non-zero rational  function  
with the independent variables $p,\,x\in \R ^n$ and hence
 we can take
a special matrix
$\al =\al _1 =- H_1\Lambda _1  H_1^{-1}$, 
whose elements are all rational  functions of
 $p,\,x\in \R ^n$. Further 
we can obtain non-zero rational  function solutions
by (\ref{dtp}).

\noindent (1.2) Let $\mu _i,\ 1\le i\le N_1$, be distinct real numbers,
$f_i:\R ^n\to \R$, $1\le i\le N_2,$ be non-zero  
rational  functions,
and $g_i:\R ^n\to \R$, $1\le i\le N_2,$ be non-zero distinct 
rational  functions so that $\prod _{1\le i\le N_1,1\le j\le N_2}(\mu _i-g_j)
\mbox{ $ / \hspace{-0.9em}\equiv $ }0$.
It follows from (\ref{deth2}) that in this case,
 $\textrm{det}H_2$ 
is a non-zero rational  function 
with the independent variables $p,\,x\in \R ^n$ and then
 we can choose
a special matrix
$\al =\al _2 =- H_2\Lambda _2  H_2^{-1}$, 
whose elements are all rational  functions of $p,\,x\in \R ^n$. In this way, 
we can obtain a class of non-zero rational  function solutions
by (\ref{dtp}).

{\bf (2) Analytic function solutions:}

\noindent (2.1) Let $\gamma _1,\cdots ,\gamma _N$ be real numbers and 
$h_i,\ 1\le i\le N$, be any analytic functions
to satisfy $|h_i(y)|\le B,\ 1\le i\le N, \ y\in \R^n$. There are  a lot
of functions of this kind. For example, $\textrm{sin}q(y), \textrm{cos}q(y),
\textrm{tanh}q(y),\textrm{sech}q(y)$, where $q:\R ^n\to \R$ is any analytic 
function. We choose 
\[ f_i=f_i(a(\la _i)\cdot p +x)=h_i(a(\la _i)\cdot p +x)+\gamma _i,\ 1\le i\le N.
\]
This moment, $\textrm{det}H_1$
 hasn't zero points with respect to   $(p,x)\in \R ^{2n}$ while
\begin{equation}|\gamma  _i|>B ,\ 1\le i\le N,\ 
 |\gamma _i-\gamma _j|>2B ,\ 1\le i< j\le N.
\label{gammacond}
\end{equation}
Therefore under the condition (\ref{gammacond}) 
we may take a special matrix
$\al =\al _1 =- H_1\Lambda _1  H_1^{-1}$, whose elements are all analytic
functions of
 $p,\,x\in \R ^n$. This can result in  
nonzero analytic function solutions
by (\ref{dtp}).

\noindent (2.2) Let $\mu _i,\ 1\le i\le N_1$,
be distinct real numbers,
$f_i:\R^n\to \R,\ 1\le i\le N_2$, 
be non-negative or
non-positive analytic functions, and 
$h_i:\R^n\to \R,\ 1\le i\le N_2$, be
analytic functions to satisfy $|h_i(y)|\le B$, $1\le i\le N_2$. 
We choose 
\[g_i=g_i(a(\la _i)\cdot p +x)=h_i(a(\la _i)\cdot p +x)+\gamma _i,
\ 1\le i\le N_2,\]
where $\gamma _i,\ 1\le i\le N_2$,
are all real numbers so that 
\[ |\gamma _i-\gamma _j| >2B,\ 1\le i< j\le N_2,\ |\gamma _i|>B+
\max_{1\le i\le N_1} |\mu _i |,\ 1\le i\le N_2.\]
In this way,  $\textrm{det}H_2$
 hasn't zero points with respect to   $(p,x)\in \R ^{2n}$.
Therefore 
we can take a special matrix
$\al =\al _2 =- H_2\Lambda _2  H_2^{-1}$, whose elements are all analytic
functions of
 $p,\,x\in \R ^n$. This may yield a class of   
nonzero analytic function solutions
by means of (\ref{dtp}).
\label{solutionsection}

\section{Discussions and remarks}
\setcounter{equation}{0}
From the mathematical point of view, it is very interesting  
to  generate  more general integrable systems. Such an example
may be introduced for 
the Lax integrable system in Section 2. We display that more general
system here
\begin{eqnarray}&&
\sum_{\ba {c}\vspace{-7mm}\\ {\scriptstyle k+l=m}\vspace{-2mm} 
\\ {\scriptstyle -M\le k,l\le M}
\ea }[A_{ik},A_{jl}]
+\sum_{\ba {c}\vspace{-7mm}\\ {\scriptstyle k+l=m}\vspace{-2mm}
\\{\scriptstyle -M\le k,l\le M}
\ea }\Bigl(
a_{jk}\frac {\part A_{il}}{\part x_j}-a_{ik}\frac {\part A_{jl}}{\part x_i}
\Bigr)=0,\nonumber\\&& 
\qquad\qquad\qquad\qquad\qquad\qquad\qquad
-2M\le m\le -M-1,\ M+1\le m\le 2M,\nonumber\\&&
\sum_{\ba {c}\vspace{-7mm}\\ {\scriptstyle k+l=m}\vspace{-2mm}
\\{\scriptstyle -M\le k,l\le M}
\ea }[A_{ik},A_{jl}]
-\frac{\part A_{im}}{\part p_j}+\frac{\part A_{jm}}{\part p_i}
+\sum_{\ba {c}\vspace{-7mm}\\ 
{\scriptstyle  k+l=m}\vspace{-2mm}\\{\scriptstyle -M\le k,l\le M}
\ea }\Bigl(
a_{jk}\frac {\part A_{il}}{\part x_j}-a_{ik}\frac {\part A_{jl}}{\part x_i}
\Bigr)=0,\nonumber \\ &&
\qquad\qquad\qquad\qquad\qquad\qquad\qquad\qquad\qquad\qquad\qquad\qquad
 -M\le m\le M,\qquad\qquad\nonumber
\end{eqnarray}
where $1\le i,j\le n$.
It is in agreement with the compatibility conditions of the following 
spectral problems with negative powers of the spectral parameter
\[ \Bigl[\Bigl(\frac {\part }{\part p_i}-\Bigl(\sum_{k=-M}^Ma_{ik}\la ^k\Bigr)
\frac {\part }
{\part x_i}\Bigr)\Bigr]\Psi =-\Bigl(\sum_{l=-M}^MA_{il}\la ^l\Bigr)
\Psi,\ 1\le i\le n,\] 
where $M\ge 0$.
A more general Lax set  of spectral problems
\[\Bigl[\Bigl(\frac {\part }{\part p_i}-\Bigl(\sum_{k=-M'}
^{M''}a_{ik}\la ^k\Bigr)
\frac {\part }
{\part x_i}\Bigr)\Bigr]\Psi =-\Bigl(\sum_{l=-M'}^{M''}A_{il}\la ^l\Bigr)
\Psi,\ 1\le i\le n, \]
where $M',\, M''\ge 0$,
may arise as a reduction of this Lax set with $M=\textrm{max}(M',M'')$
and vice versa.
The Lax integrable system displayed above includes  the generalized 
self-dual Yang-Mills flows in Ref. \cite{Gu2}. It seems more reasonable 
that it is considered as a ``universal'' integrable system, whereas 
 it is too general
to lose some concrete characteristics. It would be valuable to know whether
its more reductions have physical interpretations.
We may also construct 
a class of generalized chiral field equations similar to one in Ref. 
\cite{Ward2}.
Essentially the deduction is the same but we should note particular
properties. 

We remark
that there has been a huge class of higher-dimensional nonlinear
integrable equations which can be solved through the nonlocal Riemann-Hilbert
method proposed by Zakharov and Manakov \cite{ZakharovM}. 
This kind of 
equations are connected with the Lax representations defined by
higher-order differential operators with respect
to some independent variables. However, our initial 
Lax integrable system (\ref{laxis1}--\ref{laxis4}) 
is derived by the spectral problems involving higher-order powers of the 
parameter $\la $ other than higher-order differential operators. 
Therefore 
%the methods in this letter and in \cite{ZakharovM} are different 
%from each other and 
there isn't a direct relation between two classes of resulting integrable 
equations 
although there exist 
the same reductions of them,
for instance, KP equation. On the other hand,
our explicit solutions presented in Section \ref{solutionsection} are quite
broad because many arbitrary functions are involved in the construction of
solutions and thus it is difficult to discuss their properties. 
But we should be able to 
study some important properties of sub-classes of the obtained solutions,
for example, 
localization property, the existence of the instanton like solutions and 
the soliton interactions etc. as in the work by Zhou 
\cite{Zhou}. In particular, some careful consideration
about specific reductions
deserves a further investigation.

\vskip 5mm

\noindent{\bf Acknowledgments:} The author
would like to thank Prof. B. Fuchssteiner for his kind hospitality at 
Automath Institute, Paderborn University, Germany and Dr. W.  Oevel 
 for valuable discussions.
The author  is also indebted to Prof. C. H. Gu 
for his report on Darboux transformations at Paderborn University
and  to the referee 
 for helpful suggestions.
This work was supported by 
the National Natural Science Foundation of China,
the Alexander von Humboldt Foundation
and the Project `Venus'(Qi-Ming-Xing) of the
Shanghai Science and Technology Commission.

\end{document}